\documentclass[preprint,showpacs,preprintnumbers,amsmath,aps]{revtex4}
\usepackage{graphics}

\def\be{\begin{equation}}
\def\ee{\end{equation}}
\def\bea{\begin{eqnarray}}
\def\eea{\end{eqnarray}}
\def\bse{\begin{subequations}}
\def\ese{\end{subequations}}
\def\bma{\begin{mathletters}}
\def\ema{\end{mathletters}}
\def\C{\hbox{$\mit I$\kern-.6em$\mit C$}}

\begin{document}

\title{Entanglement in the scattering process by local impurity}

\author{Dong Yang}
\affiliation{Department of Modern Physics, University of Science and Technology of
China, Hefei, Anhui 230026, People's Republic of China} \affiliation{Zhejiang Institute
of Modern Physics, Zhejiang University, Hangzhou, Zhejiang 310027, People's Republic of
China}

\author{Shi-Jian Gu}
\affiliation{Department of Physics, The Chinese University of Hong Kong, Hong Kong,
China}
\affiliation{Zhejiang Institute of Modern Physics, Zhejiang University,
Hangzhou, Zhejiang 310027, People's Republic of China}

\author{Haibin Li}
\affiliation{Zhejiang Institute of Modern Physics, Zhejiang University, Hangzhou,
Zhejiang 310027, People's Republic of China}

\date{\today}

\begin{abstract}
We study entanglement in the scattering processes by fixed impurity and Kondo impurity. The fixed impurity plays a
role as spin state filter that is employed to concentrate entanglement between the scattering particle and the
unscattering particle. One Kondo impurity can entangle two noninteracting scattering particles while one
scattering particle can entangle two separate noninteracting Kondo impurities.
\end{abstract}

\pacs{03.67.-a}

\maketitle

\section{Introduction}
Entanglement lies at the heart of quantum information and quantum computation \cite{Nielsen1}. It is responsible for
most of quantum phenomena, such as quantum teleportation, dense coding, quantum cryptography \cite{Bennett}.
Now it is regarded as a kind of useful resource in quantum information process. Much efforts were devoted to
the entanglement capacity of unitary evolution and interaction Hamiltonian. For exmaple, Childs et al.
\cite{Childs} found an explicit formula for the maximum entanglement created by a class of two-qubit Hamiltonians,
including the Ising interaction and the anisotropic Heisenberg interaction. Nielsen et al \cite{Nielsen2}
developed a theory quantifying the strength of quantum dynamical operations. Experimentally, entanglement creation
and distillation are of importance \cite{Bennett1}. Currently, these tasks are mainly realized by optics. In this paper,
 We study the scattering process by the $\delta$ interaction created by the fixed impurity and Kondo impurity and show
that entanglement distillation and creation can be realized in the scattering process.

The paper is structured as follows. Section \ref{se:filter} discusses the scattering process with fixed impurity
and introduces spin state filter that is later employed to distill entanglement. Section \ref{se:Kondo} considers
the scattering process with Kondo impurity that is utilized to create entanglement. Section \ref{se:sum} concludes
with a summary.

\section{Spin State Filter}
\label{se:filter}
In this section, we discussed the scattering process that a free particle with spin is scattered by an
impurity with fixed spin. We show that the scattering process plays a role as spin state filter, then we utilize
the filter to concentrate entanglement between two non-maximally entangled particles.

First let us consider the problem of the free particle scattering with the $\delta$ potential. The Hamiltonian is
\be
H=-\frac{1}{2}\frac{d^{2}}{dx^{2}}+r\delta(x),
\ee
where we set $\hbar=m=1$ and $r$ is the strength of the $\delta$ potential. The $\delta$ potential comes from a
local impurity without a spin. Suppose the incident wave function
is $\phi_{i}=e^{ikx}$, then the solution of the scattering process is
\be
\phi_{f}=\left\{\begin{array}{lr}e^{ikx}+Re^{-ikx},&x<0,\\Se^{ikx},&x>0,\end{array}\right.
\ee
where $S=1/(1+i\xi)$ is the transmission amplitude, $R=S-1$ the refraction amplitude, $\xi=r/k$.

Now suppose a free particle with spin is scattered by a impurity with fixed spin state. The situation occurs where
an free electron is scattered by a localized electron whose spin is fixed, say spin up. If the electric interaction is
neglected, the scattering process can be described by the Toy model,
\be
H=-\frac{1}{2}\frac{d^{2}}{dx^{2}}+r\delta(x)(1-\sigma_{z}),
\ee
where $\sigma_{z}$ is the Pauli operator. In the eigenbasis of $\sigma_{z}$, the potential is written as
$2r\delta(x)\left(\begin{array}{cc}0&0\\0&1\end{array}\right)$. That is to say the spin of the particle is visible
to the impurity on the origin. If the particle is spin-up, the impurity lets the particle go through freely, just
as the impurity is nonexisting. If the particle is spin-down, the impurity acts as the $2r\delta(x)$ potential. For
the general case, where the particle lies in the coherent state of spin-up and spin-down, The problem can be
solved in the spirit of fraction wave method: the potential is diagonalised and the eigenvalues and eigenstates
are found; for each of the eigenstates, the scattering problem is solved for the potential determined
by its corresponding eigenvalue; the incident wave state is expanded in the eigenstates and the scattered wave
state is the superposition of the scattered wave states of the eigenstates. Now suppose the incident wave
state is expanded as
\be
\phi_{i}=e^{ikx}(\alpha|0\rangle+\beta|1\rangle),
\ee
where $|0\rangle$ is spin-up and $|1\rangle$ spin-down and $|\alpha|^{2}+|\beta|^{2}=1$. The solution can be obtained,
\be
\phi_{f}=\left\{\begin{array}{lr}\alpha e^{ikx}|0\rangle+\beta (e^{ikx}+Re^{-ikx})|1\rangle,& x<0,\\e^{ikx}
(\alpha|0\rangle+\beta S|1\rangle),&x>0\end{array}\right.
\ee
where $S=1/(1+i\xi)$, $\xi=2r/k$. Note that in the transmitted part, the amplitude of the spin-down is decreased.
The effect of the scattering process on the spin state is like a spin filter.

Spin filter can be employed to distill entanglement between two non-maximally entangled particles \cite{Gisin}.
Suppose the state of two particles is
\be
\phi_{i}=e^{-ikx_{2}}e^{ikx_{1}}(a|00\rangle+b|11\rangle), a<b,
\ee
in which $|a|^{2}+|b|^{2}=1$. The entanglement of pure bipartite state is measured by von Neumann entropy of the
reduced state of any particle, $E=-tr\rho_{1}log\rho_{1}=-|a|^{2}log|a|^{2}-|b|^{2}log|b|^{2}$, where $\rho_{1}
=|a|^{2}|0\rangle\langle0|+|b|^{2}|1\rangle\langle1|$. Suppose particle one is scattered by the impurity. According
to the fraction wave method,
\be
\begin{array}{l}
e^{ikx_{1}}|0\rangle\rightarrow\left\{\begin{array}{lr}e^{ikx_{1}}|0\rangle,&x<0,\\e^{ikx_{1}}|0\rangle,&x>0,
\end{array}\right.\\
e^{ikx_{1}}|1\rangle\rightarrow\left\{\begin{array}{lr}(e^{ikx_{1}}+Re^{-ikx_{1}})|1\rangle,&x<0,\\
Se^{ikx_{1}}|1\rangle,&x>0.\end{array}\right.
\end{array}
\ee
After the scattering process, the two particle state is of the form,
\be
\phi_{f}=\left\{\begin{array}{lr}ae^{-ikx_{2}}e^{ikx_{1}}|00\rangle+be^{-ikx_{2}}(e^{ikx_{1}}+Re^{-ikx_{1}})
|11\rangle,&x<0,\\
e^{-ikx_{2}}e^{ikx_{1}}(a|00\rangle+bS|11\rangle),&x>0.\end{array}\right.
\ee
We concern mainly about the transmitted part. When $|a|=|bS|$, the transmitted state is maximally entangled.
Given $a,k$, we can modify the strength $r$ to satisfy $|a|=|bS|$. Actually, we can vary the direction of the
impurity to achieve the same effect in a simpler way. Notice that the probability to obtain a maximally entangled
state is less than the maximum one $2|a|^{2}$ that can be obtained theoretically. The reason lies at the refracted
part still contains entanglement as the refracted term is not zero. Indeed, there exists the optimal direction
of the impurity that the probability to obtain maximally entangled state is maximum by the scattering process.\\

\section{Creating Entanglement}
\label{se:Kondo}
In this section, we introduce the scattering process that a free particle are scattered by Kondo impurity. Then
we show how to utilize the process to create entanglement between two noninteracting particles and how to create
entanglement between two noninteracting Kondo impurities. A complementary relation is given between the strength of the
$\delta$ potential and the rate of entanglement creation.

The impurity is called Kondo impurity if the spin of the impurity is free and interacts with the particle. The
scattering process is described by the Kondo model:
\be
H=-\frac{1}{2}\frac{d^{2}}{dx^{2}}+r\delta(x)\vec{\sigma_{1}}\cdot\vec{\sigma_{0}},
\ee
where $\vec{\sigma_{1}}$ is the spin vector operator of the particle and $\vec{\sigma_{0}}$ that of the Kondo
impurity. The eigenbasis of $\vec{\sigma_{1}}\cdot\vec{\sigma_{0}}$ is,
\bea
|\lambda_{1}\rangle&=&|00\rangle,\nonumber\\
|\lambda_{2}\rangle&=&|11\rangle,\nonumber\\
|\lambda_{3}\rangle&=&\frac{1}{\sqrt{2}}(|01\rangle+|10\rangle),\nonumber\\
|\lambda_{4}\rangle&=&\frac{1}{\sqrt{2}}(|01\rangle-|10\rangle),
\eea
where $\lambda_{i}$ are the corresponding eigenvalues, $\lambda_{1}=\lambda_{2}=1, \lambda_{3}=-2,
\lambda_{4}=0$. In the eigenbasis, $H$ is expressed as
\be
H=-\frac{1}{2}\frac{d^{2}}{dx^{2}}+r\delta(x)\sum_{i}\lambda_{i}|\lambda_{i}\rangle\langle\lambda_{i}|
\ee
Solving the scattering process when the incident wave state is $\phi_{i}=e^{ikx}|\lambda_{i}\rangle$ gives,
\be
\phi_{f}=\left\{\begin{array}{lr}(e^{ikx}+R_{i}e^{-ikx})|\lambda_{i}\rangle,&x<0,\\
S_{i}e^{ikx}|\lambda_{i}\rangle,&x>0,\end{array}\right.
\ee
where $S_{i}=1/(1+i\xi_{i})$, $\xi_{i}=r\lambda_{i}/k$.

Generally the incident wave state is
\be
\phi_{i}=e^{ikx}|\chi\rangle=e^{ikx}\sum_{i}c_{i}|\lambda_{i}\rangle,
\ee
and the scattering state is
\be
\phi_{f}=\left\{\begin{array}{lr}\sum_{i}c_{i}(e^{ikx}+R_{i}e^{-ikx})|\lambda_{i}\rangle,&x<0,\\
e^{ikx}\sum_{i}c_{i}S_{i}|\lambda_{i}\rangle,&x>0,\end{array}\right.
\ee

After the scattering process $\phi_{f}$ is projected onto two subspaces: one is $x<0$, and the other is $x>0$.
The transmitted part $x>0$ is retained while that $x<0$ is discarded. In other words, we only concern about
the scattering term.
Notice that in the scattering term the wave function of coordinate is the same as the incident wave. Because we
pay attention to entanglement between the spins, we omit the wave function of coordinate and rewrite the effect
on the spins as following,
\bea
|00\rangle&\rightarrow& S_{1}|00\rangle,\nonumber\\
|11\rangle&\rightarrow& S_{2}|11\rangle,\nonumber\\
|01\rangle&\rightarrow&\frac{S_{3}+S_{4}}{2}|01\rangle+\frac{S_{3}-S_{4}}{2}|10\rangle,\nonumber\\
|10\rangle&\rightarrow&\frac{S_{3}-S_{4}}{2}|01\rangle-\frac{S_{3}+S_{4}}{2}|10\rangle.
\eea
Suppose initially the impurity is polarized in spin-up and the particle is in a general state,
\be
(\alpha|0\rangle+\beta|1\rangle)|0\rangle\rightarrow\alpha
S_{1}|00\rangle+\frac{\beta(S_{3}-S_{4})}{2}|01\rangle-\frac{\beta(S_{3}+S_{4})}{2}|10\rangle.
\ee
Once the particle is transmitted, the spin state of the particle is entangled with that of the Kondo impurity.
 It appears that the Kondo impurity does not
play a role as spin state filter. However, Kondo impurity combining with additional
measurement on it would have the same effect as spin state filter. Here we use it to
concerntrate entanglement as the fixed impurity. Of course, extra measurement on the
Kondo impurity is required after scattering. As the case discussed in the fixed
impurity, \be
(a|00\rangle+b|11\rangle)|0\rangle\rightarrow\left(aS_{1}|00\rangle-\frac{b(S_{3}+S_{4})}{2}|11\rangle\right)|0\rangle
+\frac{b(S_{3}-S_{4})}{2}|10\rangle|1\rangle. \ee If
$|aS_{1}|=|\frac{b(S_{3}+S_{4})}{2}|$, then the projection onto $|0\rangle\langle0|$
will give the maximally entangled state. Notice that different measurements can be
performed on the Kondo impurity. The capacity of the concentrating entanglement of the
two kinds of impurities can be compared.

An important property of Kondo impurity is that it can be employed to create entanglement between two
noninteracting scattering particles. The process is as following. Suppose initially the Kondo impurity $P_{0}$ is
polarized in
$|1\rangle_{0}$, the two noninteracting particles in product state $|0\rangle_{2}|0\rangle_{1}$. First $P_{1}$ is
scattered by the Kondo impurity. If $P_{1}$ is transmitted, then $P_{2}$ is scattered by the impurity.
If $P_{1}$ is refracted,
the impurity $P_{0}$ is polarized again in $|1\rangle_{0}$ and a new $P_{1}$ is incident on the impurity.
In all, we just
consider the possibility when $P_{1}$ and $P_{2}$ are scattered with the same impurity $P_{0}$ one by one and
both particles are transmitted. The evolution is
\bea
|0\rangle_{2}|0\rangle_{1}|1\rangle_{0}\rightarrow|0\rangle_{2}\left(\frac{S_{3}^{1}
+S_{4}^{1}}{2}|0\rangle_{1}|1\rangle_{0}+\frac{S_{3}^{1}-S_{4}^{1}}{2}|1\rangle_{1}|0\rangle_{0}\right)\nonumber\\
\rightarrow\frac{S_{3}^{2}+S_{4}^{2}}{2}\frac{S_{3}^{1}+S_{4}^{1}}{2}|0\rangle_{2}|0\rangle_{1}|1\rangle_{0}
+\left(\frac{S_{3}^{2}-S_{4}^{2}}{2}\frac{S_{3}^{1}+S_{4}^{1}}{2}|1\rangle_{2}|0\rangle_{1}
+\frac{S_{1}^{2}(S_{3}^{1}-S_{4}^{1})}{2}|0\rangle_{2}|1\rangle_{1}\right)|0\rangle_{0},
\eea where the first arrow denotes the $P_{1}$ is scattered and the second arrow
denotes the resulted transmitted state when both particles are transmitted. Measurement
on the impurity is performed in the basis $\{|0\rangle,|1\rangle\}$. Once the outcome
$|0\rangle$ occurs, $P_{1}$ is entangled with $P_{2}$. Here we remark that the initial
spin polarized directions are of importance. Explicitly, when the initial state of the
three particles is $|0\rangle_{2}|0\rangle_{1}|0\rangle_{0}$, no entanglement will
exist in the final transmitted term.

Exchanging the roles of the scattering particle and the impurity, a moving particle can be used to create
entanglement between two separate noninteracting impurities. Suppose $P_{1},P_{2}$ are two Kondo impurities located
at $-a$ and $a$ respectively. A particle $P_{0}$ moves from left to right scattering with $P_{1}$ and $P_{2}$ sequently.
The evolution is described by the Hamiltonian,
\be
H=-\frac{1}{2}\frac{d^{2}}{dx^{2}}+r_{1}\delta(x+a)\vec{\sigma}\cdot\vec{\sigma_{1}}
+r_{2}\delta(x-a)\vec{\sigma}\cdot\vec{\sigma_{2}}.
\ee
Strictly speaking, we should solve the multipartite scattering problem. Here we mainly consider the first order
scattering that means the contributions that refract to and fro between the impurities is
neglected. The transmitted part $x>a$ is mainly from the direct transmission. Suppose the initial state is
$|1\rangle_{0}|0\rangle_{1}|0\rangle_{2}$, the evolution is
\bea
|1\rangle_{0}|0\rangle_{1}|0\rangle_{2}\rightarrow\left(\frac{S_{3}^{1}
-S_{4}^{1}}{2}|0\rangle_{0}|1\rangle_{1}-\frac{S_{3}^{1}+S_{4}^{1}}{2}|1\rangle_{0}|0\rangle_{1}\right)|0\rangle_{2}\nonumber\\
\rightarrow\frac{S_{3}^{1}+S_{4}^{1}}{2}\frac{S_{3}^{2}+S_{4}^{2}}{2}|1\rangle_{0}|0\rangle_{1}|0\rangle_{2}
+|0\rangle_{0}\left(\frac{S_{1}^{2}(S_{3}^{1}-S_{4}^{1})}{2}|1\rangle_{1}|0\rangle_{2}-
\frac{S_{3}^{1}+S_{4}^{1}}{2}\frac{S_{3}^{2}-S_{4}^{2}}{2}|0\rangle_{1}|1\rangle_{2}
\right), \eea Measurement on the moving particle is performed in the basis
$\{|0\rangle,|1\rangle\}$. The occurrence of the outcome $|0\rangle$ will reduce the
two impurities to entangled state.

\section{Summary}
\label{se:sum}
In this article, we showed the scattering process can be employed to tackle with entanglement problem. Spin state
filter can be realized by scattering with fixed impurity. Entanglement between two noninteracting particles can
be created by scattering with the Kondo impurity. On the contrary, entanglement between two noninteracting Kondo
impurities can be created by scattering with the same particle. In the case of entanglement creation. we just
considered the scattering process with the first order. Strictly speaking, the scattering process with high
orders should be included and deserves further investigation.


\end{document}